\documentclass{Interspeech}

\usepackage[table,xcdraw]{xcolor}
\usepackage{svg}
\usepackage{amsmath}
\usepackage{multirow}
\usepackage{subfig}
\usepackage{floatrow}
\usepackage{tikz}
\DeclareUnicodeCharacter{2212}{$-$}

\usepackage[tableposition=top]{caption}

\interspeechcameraready

\title{On the influence of language similarity in non-target speaker verification trials}

\author[affiliation={1}]{Paul M.}{Reuter}
\author[affiliation={2}]{Michael}{Jessen}

\affiliation{Fraunhofer Institute for Digital Media Technology}{Division Hearing, Speech and Audio Technology}{Oldenburg, Germany}
\affiliation{Department of Text, Speech and Audio}{Bundeskriminalamt}{Wiesbaden, Germany}
\email{paul.maria.reuter@idmt.fraunhofer.de, michael.jessen@bka.bund.de}

\usepackage{comment}

\begin{document}

\floatsetup[figure]{style=plain,subcapbesideposition=top}
\floatsetup[table]{capposition=top}

\maketitle

\begin{abstract}

In this paper, we investigate the influence of language similarity in cross-lingual non-target speaker verification trials using a state-of-the-art speaker verification system, ECAPA-TDNN, trained on multilingual and monolingual variants of the VoxCeleb dataset. Our analysis of the score distribution patterns on multilingual Globalphone and LDC~CTS reveals a clustering effect in speaker comparisons involving a training language, whereby the choice of comparison language only minimally impacts scores. Conversely, we observe a language similarity effect in trials involving languages not included in the training set of the speaker verification system, with scores correlating with language similarity measured by a language classification system, especially when using multilingual training data. 


\end{abstract}

\section{Introduction}





There is continued interest over the years in how and to what extent automatic speaker recognition is affected by language. A common research paradigm to address this question has been to study mismatch between the language spoken by the questioned speaker and known speaker~\cite{brummer07_fusion, MISRA201858, Nechansky22_impact, thienpondt22_tackling}. Some researchers approached the language effect by studying language mismatch not between questioned and known speaker but between the language spoken by both of them (test language) and the data used for embedding-level training \cite{matejka17_interspeech, BAHMANINEZHAD20217, sztaho23}, for score normalisation \cite{matejka17_interspeech, Skarnitzl_Asiaee_Nourbakhsh_2020}, or calibration \cite{sztaho23, van2017experiments}.
Most of these studies focus on overall performance characteristics such as EER or Cllr and some of them show how the performance loss due to language mismatch relative to language match could be mitigated.
Forensic automatic speaker recognition, the main target of this work, has an interest not only in overall speaker discrimination performance, but also in how the results of a system test (validation) are distributed in Tippett plots or other distribution graphs~\cite{drygajlo15_methodological, MORRISON2021299}. The current study focuses entirely on different-speaker comparisons and on the score distributions that these comparisons generate. Some of these comparisons involve the same language (language match) others different languages (language mismatch) from a set of languages available in the corpora to be used here. We hypothesise that different-speaker comparisons involving the same language yield higher scores than those involving different languages (H1). We also hypothesise that among the different-language (aka cross-lingual) comparisons (aka trials), those that involve languages that are in some way similar to each other yield higher scores than those that involve languages that are more distant to each other (H2). What counts as similar or distant will be operationalized by using a language classification system. Moreover, it will be examined to what extent the composition of the deep-level training data influences the results, by using either multi-lingual or mono-lingual training sets. The novelty of this research lies in its focus on the language-induced score shift patterns and the range of factors considered for their explanation. It has implications for forensics in particular, but also more generally, including the current discussion on potential bias in AI processing. The score distribution of different-speaker trials is important for score normalisation methods such as t-norm or s-norm~(\cite{MORRISON201937} for a summary of forensically-oriented studies using these normalisation methods; see also~\cite{BAHMANINEZHAD20217}). It also has relevance for calibration. When no data are available to form a normalization or calibration set that is equilingual to the recordings in the forensic comparison, it is important to know to what extent the normalisation or calibration procedure would be biased when using a non-matching language set and whether the effect can be predicted by language distance or other factors.


\section{Methods}

\subsection{Language classification}

\begin{figure}[htb]
  \centering
  \scalebox{0.8}{
  \includesvg[width=0.6\linewidth]{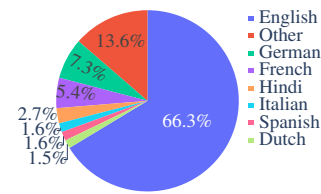}}
  \caption{Language distribution in VoxCeleb1 and 2}
  \label{fig:voxceleb_languages}
\end{figure}

For language classification, a ResNet34~\cite{he16_resnet} was trained on VoxLingua107 \cite{valk21_voxlingua} using cross-entropy loss. The model was used to analyze the distribution of languages spoken in the VoxCeleb1 and 2 datasets \cite{nagrani17_interspeech, chung18b_interspeech} which served as training data for the speaker verification models (see Fig. \ref{fig:voxceleb_languages}).

\subsection{Speaker verification}

ECAPA-TDNN~\cite{desplanques20_interspeech} was adopted as the speaker verification model architecture as it is commonly used in state-of-the-art speaker verification~\cite{thienpondt21_interspeech, zhao23_pcf, chen23m_interspeech}. It is built upon the original x-vector architecture~\cite{snyder18_xvector} and employs several enhancements regarding channel attention, propagation and aggregation. 
In our experiments, we used two models of the same ECAPA-TDNN architecture trained on different datasets. The first model is the pretrained ECAPA-TDNN model from Speechbrain~\cite{ravanelli2021speechbraingeneralpurposespeechtoolkit} trained on VoxCeleb1 and 2 (also referred to as system Vox1+2). The second model was trained exclusively on English utterances from VoxCeleb2 (also referred to as system Vox2-en). Both models were trained with the AAM-Softmax loss~\cite{deng19_arcface} for 15 epochs, following the Speechbrain recipe, resulting in an EER of 0.90\,\% and 1.49\,\% on the VoxCeleb1-O test set, respectively.

\subsection{Datasets and preparation}

We conducted our experiments on two multilingual datasets.

\subsubsection{Globalphone}

Globalphone~\cite{schultz02_icslp} is a multilingual database consisting of high-quality recordings of read sentences. It is uniform across languages with respect to the recording conditions and the collection scenario. We performed voice activity detection on every file to remove non-speech segments and cut the remaining speech content to 5 seconds length. From every available language, 40 male and/or female speakers were randomly selected with two recordings per speaker. Since English is the main language in VoxCeleb but not contained in Globalphone, we included 40 random male and female speakers from Librispeech~\cite{panayotov15_libri}, which is very similar in acoustic conditions. This left us with a total of 10 languages for male speakers and 13 languages for female speakers.

\subsubsection{LDC Multilingual CTS 2011}

LDC Multilingual CTS 2011~\cite{strassel12_odyssey} is a collection of telephone speech datasets in various languages. We performed voice activity detection and cut all recordings to 20 seconds length. Per language, 20 male and/or female speakers were randomly selected with two recordings per speaker. We ended up with a total of 16 languages for male speakers and 18 for female speakers.
\

\subsection{Comparison procedure}

For both datasets, we performed pair-wise comparisons between all possible different-speaker combinations of files within and across languages.  Speaker verification was performed using cosine similarity of the speaker embeddings. No post-processing (like score normalization) was applied. To measure the language similarity of two utterances, we use the cosine similarity between the outputs of the last hidden layer (embeddings) of the language classifier. This way, for every two-file comparison, a speaker similarity and a language similarity score were obtained.

\section{Results}

\subsection{Qualitative analysis}

\label{subsec:scores}

\begin{figure}[htb]
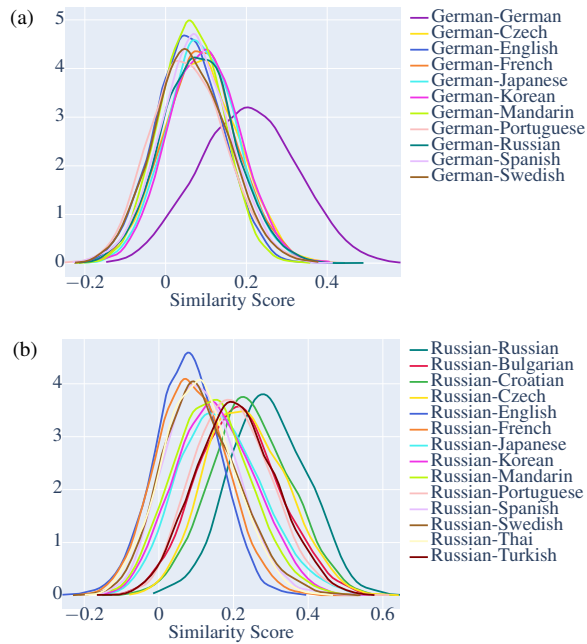

  \centering
  \sidesubfloat[]{
  \scalebox{0.8}{
 \includesvg[width=1\linewidth]{vox1+2_male_different_speaker_scores_German_updated2.svg}}} 
  \\[\baselineskip]
  \sidesubfloat[]{\scalebox{0.8}{
  \includesvg[width=1\linewidth]{vox1+2_female_different_speaker_scores_Russian_updated2.svg}}}
    \caption{Score distributions of different-speaker trials involving a) German (males) and b) Russian (females) in Globalphone set with VoxCeleb1+2 training}
    \label{fig:vox1+2_german_russian}
\end{figure}


Fig.~\ref{fig:vox1+2_german_russian} shows the score distributions of different-speaker trials in Globalphone involving German and Russian, respectively, for the system trained on VoxCeleb1+2. For both reference languages, speaker similarity scores are highest on average for the same-language case (German-German, Russian-Russian) (H1 fulfilled). However, there are noticeable differences in the distributions of different-language scores. While scores of comparisons involving German are largely unaffected by the language being compared (H2 not fulfilled), there is a shift in scores of Russian comparisons depending on the compared language with generally higher scores for related (Slavic) languages such as Bulgarian, Croatian and Czech (H2 fulfilled). Examining the average female different-speaker similarity scores of all language comparisons (see Table~\ref{table:vox1+2_mean_speaker_scores}) reveals a pattern similar to German for English and French as well as a pattern similar to Russian for other Slavic languages in the Globalphone set. We also observe higher speaker similarity scores among Mandarin, Japanese, Thai and Korean.  In the LDC set, again Slavic-Slavic comparisons resulted in higher average scores than Slavic-Non-Slavic comparisons. Increased speaker similarity scores were also found among Bengali, Punjabi, Urdu and Pashto (Indo-Iranian languages) and among Thai and Lao (Tai languages). Due to space constraints LDC results are not presented here; however, the relationship between speaker similarity and language similarity in LDC will be explored later on.


The results suggest that in cross-lingual different-speaker trials, the choice of comparison language has a small effect on speaker similarity scores if the reference language is well represented in the training dataset (such as English, German, French, see Fig.~\ref{fig:voxceleb_languages}). As part of this pattern, the score distributions of all the compared languages are tightly clustered. For reference languages not included or barely present in the training dataset (such as Slavic languages), speaker similarity scores are higher for related than for non-related comparison languages and no clustering of score distributions is observed. Cross-lingual comparisons involving a training language quite robustly produce the lowest scores (see Table~\ref{table:vox1+2_mean_speaker_scores}), presumably because the system interprets known vs known/unknown language pairs as a strong contrast, whereas unknown vs unknown language pairs are interpreted as less contrasting. 


\begin{table*}
\centering
\caption{Average female different-speaker similarity scores of language comparisons in Globalphone set with VoxCeleb1+2 training}
\label{table:vox1+2_mean_speaker_scores}
  \scalebox{0.75}{
\begin{tabular}{l|r|r|r|r|r|r|r|r|r|r|r|r|r|r}
\rowcolor[HTML]{E5ECF6} 
 & English & Swedish & French & Spanish & Portuguese & Czech & Bulgarian & Croatian & Russian & Turkish & Mandarin & Japanese & Thai & Korean  \\
 \hline
English    & 0.121 & 0.083 & 0.086 & 0.086 & 0.073 & 0.076 & 0.072 & 0.074 & 0.077 & 0.068 & 0.087 & 0.085 & 0.087 & 0.079 \\
\rowcolor[HTML]{E5ECF6} 
Swedish    &       & 0.303 & 0.080  & 0.112 & 0.111 & 0.115 & 0.095 & 0.126 & 0.114 & 0.146 & 0.131 & 0.127 & 0.148 & 0.115 \\
French     &       &       & 0.243 & 0.075 & 0.056 & 0.098 & 0.066 & 0.063 & 0.089 & 0.079 & 0.094 & 0.094 & 0.091 & 0.102 \\
\rowcolor[HTML]{E5ECF6} 
Spanish    &       &       &       & 0.274 & 0.145 & 0.080  & 0.085 & 0.117 & 0.117 & 0.118 & 0.163 & 0.152 & 0.184 & 0.153 \\
Portuguese &       &       &       &       & 0.232 & 0.167 & 0.160  & 0.193 & 0.198 & 0.170  & 0.157 & 0.141 & 0.123 & 0.132 \\
\rowcolor[HTML]{E5ECF6} 
Czech      &       &       &       &       &       & 0.297 & 0.195 & 0.205 & 0.234 & 0.154 & 0.112 & 0.146 & 0.140  & 0.125 \\
Bulgarian  &       &       &       &       &       &       & 0.214 & 0.194 & 0.214 & 0.150  & 0.09  & 0.123 & 0.079 & 0.103 \\
\rowcolor[HTML]{E5ECF6} 
Croatian   &       &       &       &       &       &       &       & 0.261 & 0.245 & 0.203 & 0.137 & 0.155 & 0.102 & 0.138 \\
Russian    &       &       &       &       &       &       &       &       & 0.293 & 0.207 & 0.142 & 0.168 & 0.106 & 0.154 \\
\rowcolor[HTML]{E5ECF6} 
Turkish    &       &       &       &       &       &       &       &       &       & 0.242 & 0.202 & 0.186 & 0.127 & 0.202 \\
Mandarin   &       &       &       &       &       &       &       &       &       &       & 0.392 & 0.284 & 0.274 & 0.298 \\
\rowcolor[HTML]{E5ECF6} 
Japanese   &       &       &       &       &       &       &       &       &       &       &       & 0.287 & 0.230  & 0.275 \\
Thai       &       &       &       &       &       &       &       &       &       &       &       &       & 0.398 & 0.204 \\
\rowcolor[HTML]{E5ECF6} 
Korean     &       &       &       &       &       &       &       &       &       &       &       &       &       & 0.338

\end{tabular}}

\end{table*}

\begin{figure}[htb]
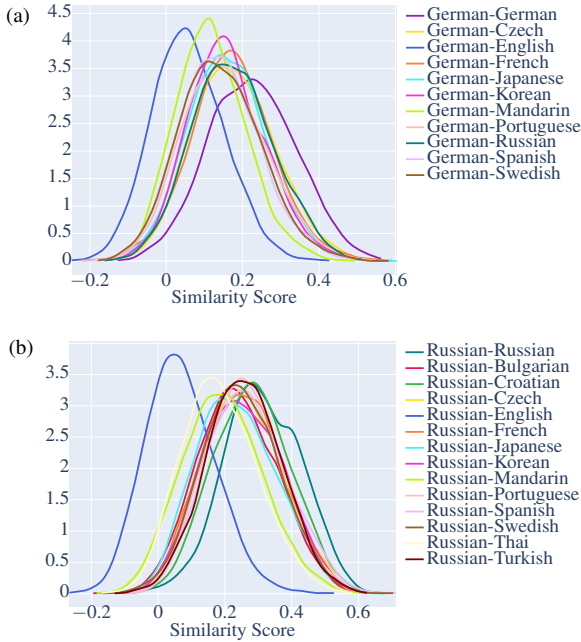

  \centering
  \sidesubfloat[]{
  \scalebox{0.8}{
  \includesvg[width=1\linewidth]{vox2-en_male_different_speaker_scores_German_updated2.svg}}} 
  \\[\baselineskip]
  \sidesubfloat[]{\scalebox{0.8}{
  \includesvg[width=1\linewidth]{vox2-en_female_different_speaker_scores_Russian_updated2.svg}}}

    \caption{Score distributions of different-speaker trials involving a) German (males) and b) Russian (females) in Globalphone set with English-only VoxCeleb2 training}
    \label{fig:vox2-en_german_russian}
\end{figure}

Score distributions of the system trained on English-only VoxCeleb2 from German and Russian comparisons are depicted in Fig.~\ref{fig:vox2-en_german_russian}. Once again, same-language comparisons achieve the highest average score (H1 fulfilled) and the system outputs lowest average scores in comparisons involving the training language (English) on Globalphone and LDC. (These two are robust patterns that re-emerge throughout this study; they will not be further mentioned beyond here.) 
Compared to system Vox1+2, neither is a clustering effect shown in the German-based comparisons of Fig.~\ref{fig:vox2-en_german_russian}\hspace{0em}a nor is a clear language similarity effect shown in Russian-based comparisons of Fig.~\ref{fig:vox2-en_german_russian}\hspace{0em}b (H2 not fulfilled). Reduction of the clustering effect may have to do with the fact that German no longer is a training language. Reduction of the language similarity effect is probably due to the fact that the system is not trained for any language other than English.

\begin{figure}[htb]
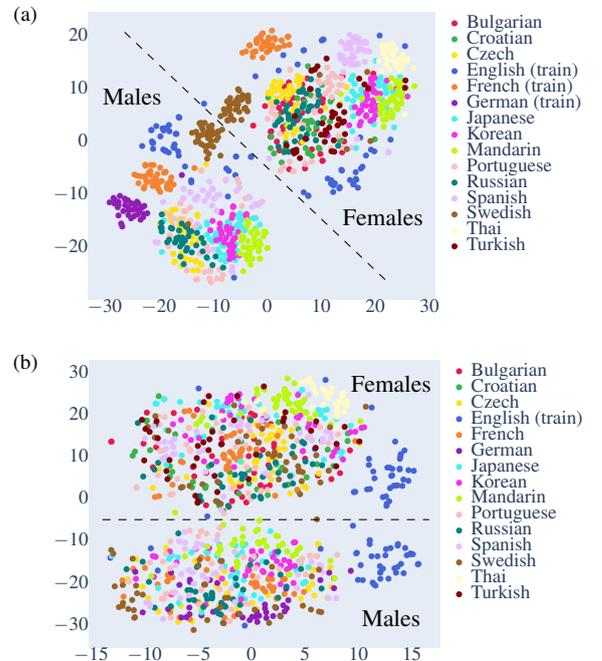

  \centering
  \sidesubfloat[]{
    \begin{tikzpicture}
      \node[anchor=south west] (img) at (0,0) {\scalebox{0.8}{\includesvg[width=1\linewidth]{vox1+2_tsne_vis_male_femal_50_auto_updated.svg}}};
      \draw[dashed] (0.9,3.85) -- (4.4,0.5);
      \node at (1., 3) {Males}; 
      \node at (4.3, 1.4) {Females}; 
    \end{tikzpicture}
  } 
  \\[\baselineskip]
  \sidesubfloat[]{
    \begin{tikzpicture}
      \node[anchor=south west] (img) at (0,0) {\scalebox{0.8}{\includesvg[width=1\linewidth]{vox2-en_tsne_vis_male_femal_50_auto_updated.svg}}};
      \draw[dashed] (0.6,2) -- (4.9,2);
      \node at (4.4, 0.7) {Males}; 
      \node at (4.4, 3.8) {Females}; 
    \end{tikzpicture}
  }
  \caption{t-SNE visualization of averaged speaker embeddings from Globalphone set with a) VoxCeleb1+2 training and b) English-only VoxCeleb2 training. The perplexity parameter was set to 50. Male and female speakers are fairly accurately separated by the drawn dashed line.}
  \label{fig:tsne_german_russian}
\end{figure}

To further understand the two different score distribution patterns in Fig.~\ref{fig:vox1+2_german_russian} and~\ref{fig:vox2-en_german_russian}, Globalphone embeddings were averaged per speaker and visualized using t-SNE~\cite{vandermaaten08a} (see Fig.~\ref{fig:tsne_german_russian}). We make two observations: Firstly, the t-SNE map is divided into male and female speakers. This is expected as biological sex affects vocal characteristics. Secondly, for the system trained on VoxCeleb1+2, speakers of languages well represented in the training dataset (English, German, French) or related languages (Swedish) form clusters in t-SNE map that are clearly separated from clusters of speakers of other languages. 
We assume that in the training of a system, speakers of the main training languages are arranged in the embedding space such that their distance to speakers of different languages is largely unaffected by the language spoken (what was seen in Fig.~\ref{fig:vox1+2_german_russian}\hspace{0em}a). As a result, speakers of training languages have no preferred association with other languages and isolate themselves in the t-SNE map unlike speakers of untrained languages. When trained on English exclusively, only English speakers form their own cluster in the t-SNE map. \\ 

\subsection{Quantitative analysis}


So far the results have shown that the effect of language similarity on non-target speaker verification scores is more complex than originally assumed when H2 was expressed. A language similarity effect tends to occur only when a system is trained on several languages and the languages that are compared are not among those the system is trained for. When instead the trained languages constitute the basis of the comparisons, the score distributions of the languages that are compared against the trained ones tend to form a cluster. To separate the effect of training in the following analysis, we split languages into the category trained (English, German and French for system Vox1+2 and English for Vox2-en) or untrained (all other languages) and consider only trained vs untrained and untrained vs untrained comparisons. In order to quantify the trends that were shown in the previous section, two procedures were used.


\begin{table}[htb]
 \caption{Score standard deviations from trained vs untrained (highlighted) and untrained vs untrained comparisons in Globalphone set for systems Vox1+2 and Vox2-en}
 \label{table:language_stds}
\centering
  \scalebox{0.9}{
\begin{tabular}{l|c|c}

 Language & Vox1+2 & Vox2-en \\
 \hline
English	 	& \cellcolor[HTML]{EBEBEB} 0.086  & \cellcolor[HTML]{EBEBEB} 0.097 \\
German		& \cellcolor[HTML]{EBEBEB} 0.090	& 0.106 \\
French		& \cellcolor[HTML]{EBEBEB} 0.097	& 0.113 \\
Swedish		& 0.099	& 0.111\\
Spanish		& 0.107	& 0.112\\
Portuguese	& 0.109	& 0.112\\
Czech		& 0.112	& 0.112 \\
Bulgarian	& 0.116	& 0.119 \\
Croatian	& 0.113	& 0.116\\
Russian		& 0.113	& 0.115 \\
Turkish		& 0.113	& 0.115 \\
Mandarin	& 0.123 & 0.118 \\
Japanese	& 0.122	& 0.116 \\
Thai		& 0.121	& 0.120  \\
Korean		& 0.125	& 0.117\\
\end{tabular}}
\end{table}

Firstly, we measured the standard deviation of cross-lingual comparison scores. The average standard deviations of male and female language sets (if available) are displayed in Table~\ref{table:language_stds}. It can be seen that for system Vox1+2 the standard deviation of scores from trained vs untrained comparisons is lower than that from untrained vs untrained comparisons. For system Vox2-en, English produces the lowest standard deviation of scores. The results show that comparisons involving a training language are less affected by the comparison language and hence have a smaller score variation which confirms the hypothesis of a clustering effect.


\begin{figure}[htb]
  \centering
  \scalebox{0.8}{
  \includesvg[width=1\linewidth]{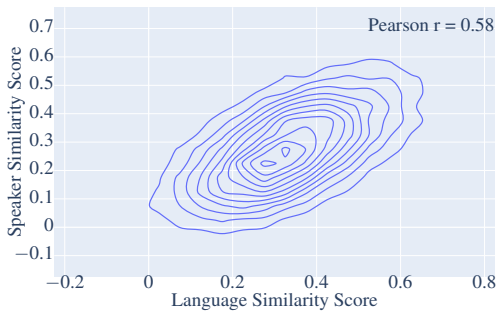}}

    \caption{Language and speaker similarity score distribution contours of male different-speaker cross-lingual trials involving Punjabi in LDC set with VoxCeleb1+2 training}
      \label{fig:corr_vox1+2}
\end{figure}


Secondly, we performed a correlation analysis between speaker similarity and language similarity. For every language considered, the Pearson correlation coefficient between speaker and language similarity scores obtained from different-speaker cross-lingual comparisons (trained vs untrained and untrained vs untrained) was calculated. In Fig.~\ref{fig:corr_vox1+2}, the speaker similarity scores of the system trained on VoxCeleb1+2 are plotted against the language similarity scores for Punjabi. There is a moderate positive correlation between speaker and language similarity. Table~\ref{table:language_correlation} displays the average of the calculated correlation coefficient from the male and female set of each language (if available).

\begin{table}[h]
 \caption{
 Pearson correlation coefficient between speaker and language similarity scores obtained from trained vs untrained (highlighted) and untrained vs untrained comparisons in Globalphone and LDC sets for systems Vox1+2 and Vox-en}

 
 \label{table:language_correlation}
\centering
  \scalebox{0.9}{
\begin{tabular}{ l|c|c||c|c  }
  & \multicolumn{2}{c||}{Globalphone} & \multicolumn{2}{c}{LDC} \\
 \cline{2-5}
 Language & Vox1+2 & Vox2-en & Vox1+2 & Vox2-en \\
 \hline
English	 	& \cellcolor[HTML]{EBEBEB} 0.076 & \cellcolor[HTML]{EBEBEB} 0.104 & \cellcolor[HTML]{EBEBEB} 0.313 & \cellcolor[HTML]{EBEBEB} 0.349 \\
German		& \cellcolor[HTML]{EBEBEB} 0.047	& 0.079 & / & / \\
French		& \cellcolor[HTML]{EBEBEB} 0.104	& 0.175 & / & / \\
Swedish		& 0.196	& 0.155 & / & /  \\
Spanish		& 0.126	& 0.175 & 0.458 & 0.433  \\
Portuguese	& 0.207	& 0.138 & / & / \\
Czech		& 0.295	& 0.123 & 0.511 & 0.358 \\
Polish      &    /   &    /   & 0.622 & 0.481 \\
Slovak      &    /   &    /   & 0.578 & 0.431 \\
Bulgarian	& 0.404	& 0.235 & / & /\\
Croatian	& 0.334	& 0.198 & / & /\\
Ukrainian   &    /   &   /    & 0.574 & 0.428 \\
Russian		& 0.359	& 0.185 & 0.549 & 0.404 \\
Turkish		& 0.243	& 0.149 & 0.482 & 0.417 \\
Mandarin	& 0.530 & 0.250 & 0.459 & 0.332 \\
Japanese	& 0.402	& 0.272 & / & / \\
Thai		& 0.454	& 0.338 & 0.581 & 0.427 \\
Korean		& 0.476	& 0.264 & / & / \\
Lao         &    /   &    /   & 0.535 & 0.435 \\
Arabic      &    /   &    /   & 0.422 & 0.348 \\
Farsi       &    /   &    /   & 0.428 & 0.320 \\
Bengali     &    /   &       & 0.518 & 0.502 \\
Punjabi     &    /   &    /   & 0.539 & 0.479 \\
Urdu        &    /   &    /   & 0.490 & 0.431 \\
Tamil       &    /   &    /   & 0.505 & 0.459 \\
\end{tabular}}

\end{table}

On both datasets system Vox1+2 shows higher correlation coefficients for languages not included in the training dataset than for languages well represented (English, German, French). This confirms the hypothesis of a language similarity effect for untrained languages. Speaker similarity scores of system Vox2-en correlate less with language similarity on average than those of system Vox1+2, presumably, because the system extracts less language information as it was only trained for English. LDC yields higher correlations than Globalphone, possibly because with LDC there are more untrained languages to compare between and a wider range of both similar and dissimilar languages.




\section{Conclusion}

In this paper, we focussed particularly on the influence of language similarity in non-target cross-lingual speaker verification trials for a state-of-the-art speaker verification system. 
Results show that in comparisons involving a training language the choice of comparison language has only a small effect on the generated scores (clustering effect). In comparisons among languages the system was not trained on, speaker similarity scores correlate with language similarity (language similarity effect). This effect is more pronounced for a system with multilingual training. Insights like these are important for forensic speaker recognition, which is characterised by sparseness of case-relevant data. Knowing what factors lie behind language-related score shifts can improve score normalisation and also calibration.



\ifinterspeechfinal
\section{Acknowledgements}

This work was partially funded by the Federal Ministry of Education and Research of Germany (BMBF) in the VIKING project (13N16239).

\else

\fi

\bibliographystyle{IEEEtran}
\bibliography{mybib}

\end{document}